\documentclass[twocolumn]{aastex61}
\usepackage{graphics,graphicx,natbib,subfigure,wrapfig}
\usepackage{ textcomp }
\usepackage{amssymb}

\newcommand{\cotwo}{CO$_{2}$}
\newcommand{\coo}{$^{13}$CO}
\newcommand{\cooo}{C$^{18}$O}
\newcommand{\coooo}{$^{13}$C$^{18}$O}

\newcommand{\hhhp}{\mbox{{\rm H}$_3^+$}}
\newcommand{\nn}{\mbox{{\rm N}$_2$}}
\newcommand{\ammonia}{\mbox{{\rm NH}$_3$}}
\newcommand{\nnhp}{\mbox{\rm N}$_{2}$\rm{H}$^+$}

\begin{document}

\title{Line Ratios Reveal \nnhp\ Emission Originates Above the Midplane in TW Hydrae}

\correspondingauthor{Kamber R. Schwarz}
\email{kschwarz@lpl.arizona.edu}

\author{Kamber R. Schwarz}
\altaffiliation{Sagan Fellow}
\affiliation{Lunar and Planetary Laboratory, University of Arizona, 1629 E. University Blvd, Tucson, AZ 85721, USA}
\author{Richard Teague}
\affiliation{Department of Astronomy, University of Michigan, 1085 South University Ave., Ann Arbor, MI 48109, USA}
\author{Edwin A. Bergin}
\affiliation{Department of Astronomy, University of Michigan, 1085 South University Ave., Ann Arbor, MI 48109, USA}

\begin{abstract}
Line ratios for different transitions of the same molecule have long been used as a probe of gas temperature. Here we use ALMA observations of the \nnhp\ J~=~1-0 and J~=~4-3 lines in the protoplanetary disk around TW Hya to derive the temperature at which these lines emit. We find an averaged temperature of 39~K with a one sigma uncertainty of 2~K for the radial range 0.8-2\arcsec, significantly warmer than the expected midplane temperature beyond 0\farcs5 in this disk. We conclude that the \nnhp\ emission in TW Hya is not emitting from near the midplane, but rather from higher in the disk, in a region likely bounded by processes such as photodissociation or chemical reprocessing of CO and \nn rather than freeze out.
\end{abstract}

\section{Introduction}
In protoplanetary disks molecular line emission is used to obtain the abundances of different species as well as measure physical properties within the disk such as temperature, turbulence, and ionization. One species which has proved useful for constraining physical properties is \nnhp, a molecular ion which emits strongly in protoplanetary disks. 
\nnhp\ is formed when \hhhp\ transfers a proton to \nn\ and is destroyed primarily by reacting with CO. Because of this, as well as the formation reaction competing with proton transfer between \hhhp\ and CO, \nnhp\ only exists at large abundances in regions with \nn\ gas but without a large CO gas abundance. As such, \nnhp\ is a potential tracer of the CO snowline \citep{Qi13b}. 

For rings of \nnhp\ emission the inner radius of the emission has been posited to trace the midplane CO snowline for TW Hya and HD 163296. For HD 163296 the CO snowline location based on \nnhp\ emission is in good agreement with the snowline location as determined by \cooo\ observations \citep{Qi15}. However, for TW Hya the snowline location of 30~au based on \nnhp\ emission is significantly farther out than midplane snowline location of 17~au derived from observations of \coooo\ \citep{Qi13b,Zhang16}. 
Additionally, \nnhp\ emission does not appear to trace the CO snowline in the V4046 Sgr disk \citep{Kastner18}.

Using physical-chemical and radiative transfer modeling, \citet{vantHoff17} argue that this discrepancy is due in part to some of the \nnhp\ emission originating from higher in the disk. The utility of \nnhp\ as a snowline tracer thus depends on the physical properties of the protoplanetary disk, including its temperature structure. 
In this letter we use the ratio of \nnhp\ J~=~4-3 to \nnhp\ J~=~1-0 emission observed in TW Hya to derive the average temperature of the \nnhp\ emitting ring. We demonstrate how temperatures derived from line ratios can inform our understanding of the likely formation pathway of a given molecule as well as what underlying conditions gave rise to the morphology of the emitting region.

\section{Observations}
The ALMA Band 3 observations targeting \nnhp\ J~=~1-0 where obtained on 2016 October 2 with 42 antennas as part of project 2016.1.00592.S (PI: K. Schwarz).
The data were calibrated and imaged using \texttt{CASA v4.7.0}. Phase and amplitude self-calibration were performed on the continuum and applied to the line spectral window. Continuum subtraction was performed using the \texttt{CASA} task \texttt{uvcontsub}.
The line data were imaged using natural weighting, resulting in a $0\farcs62$ $\times$ $0\farcs49$ beam and a per channel rms noise level of 2.2~K per 0.1~km~s$^{-1}$ channel.
Additionally, we use archival observations of the \nnhp\ J~=~4-3 transition (2011.0.00340.S, PI: C. Qi). For a detailed discussion of the data reduction we refer the reader to \citet{Qi13b}. Here we report only the properties of the final \nnhp\ image cube for comparison with the 1-0 data. The final image has a synthesized beam of  $0\farcs63$ $\times$ $0\farcs59$ and a per channel rms noise level of 0.71~K per 0.1~km~s$^{-1}$ channel.

\section{Analysis}
\begin{figure}
\setlength{\intextsep}{0pt}
\centering
    \includegraphics[width=0.5\textwidth]{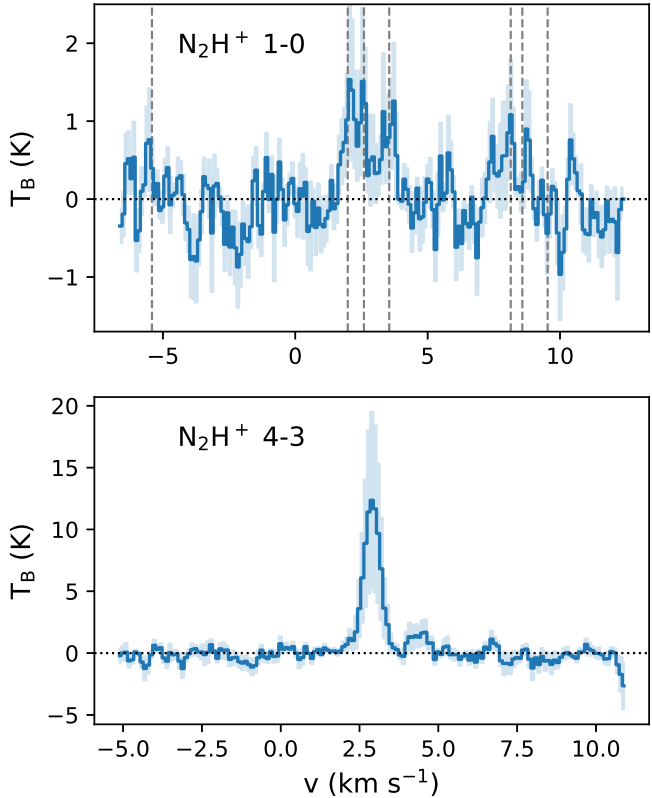}
\caption{Spectra for the \nnhp\ J = 1-0 (top) and J = 4-3 (bottom) lines after de-projection, stacking, and averaging over a ring from 0\farcs8 to 2\arcsec. Light shading indicates the $\mathrm{1\sigma}$ uncertainty in each channel. Vertical dashed lines show the expected location of the 1-0 hyperfine lines.
\label{spectra}}
\end{figure}

After standard calibration, emission from the \nnhp\ J~=~1-0 transition is not apparent in either the channel maps or moment 0 map. 
The lack of direct detection of the 1-0 transition in the image plane gives an upper limit on the integrated line intensity of 2.7~K~km~s$^{-1}$ for a single hyperfine component, where the upper limit is RMS$\times \sqrt{N_{chan}}\time \delta v$ and $\delta v = 1.5$~km~s$^{-1}$ is the line width based on the 4-3 data. The upper limit for the seven hyperfine components added in quadrature is then 7.1~K~km~s$^{-1}$.
In order to improve the signal-to-noise of the spectra we employ the stacking method for Keplerian disks introduced by \citet{Yen16}.
Using the package \texttt{eddy} \citep{eddy} the image cube is divided into concentric annuli, which are each de-projected to the system velocity assuming the physical and kinematic properties derived from previous analysis of the CS emission in TW Hya \citep{Teague18b}. The de-projected spectra are then stacked. Line emission is weakly detected in the annular bins in the range 0\farcs8 to 2\arcsec\ from source center. In comparison, bright \nnhp\ 4-3 emission is observed in the image plane in a ring from 0.8 to 1\farcs2, though weaker emission can be seen out to 2\farcs5 \citep{Qi13b}.
Figure~\ref{spectra} shows the averaged spectrum for a ring from 0.8 to 2\arcsec\ as well as the ring-averaged 4-3 spectrum for the same parameters. 

Multiple hyperfine lines are clearly seen in the averaged 1-0 spectrum. Given sufficient signal-to-noise, the relative strengths of the different hyperfine components can be used to measure the optical depth \citep[e.g.,][]{Mangum15}. 
We use the ratio of the $1_{2,3}-0_{1,2}$ line relative to the four next strongest hyperfine lines as they appear in our averaged spectrum to calculate the optical depth. For each ratio our data gives an optical depth greater than 1, ranging from 1.5 for the $1_{2,2}-0_{1,1}$ line to 66 for the $1_{2,1}-0_{1,1}$ line.
If the \nnhp\ 1-0 emission is indeed optically thick, the beam temperature should be roughly equal to the gas temperature, implying that \nnhp\ is emitting from gas colder than 2~K. As we discuss below this seems unlikely given our current understanding of the physical conditions within the TW Hya disk. A more likely explanation is that the signal-to-noise remains too poor to use the relative strength of the hyperfine components to constrain the optical depth.

The ring-averaged peak brightness temperatures of the 1-0 and 4-3 lines are $1.5\pm1$~K and $12\pm7$~K respectively, where the uncertainty is the one sigma value. \nnhp\ is posited to emit at temperatures close to the CO freeze out temperature. At the relevant radii in TW Hya the freeze out temperature is expected to be 21~K \citep[see][]{Schwarz16}. These brightness temperatures are well below this value and thus the emission is assumed to be optically thin.
Radial variations in the intensity of the 4-3 line within our annular bins likely contribute to the uncertainty for the ring-averaged peak brightness. 
However, as we are directly comparing the 4-3 data to the 1-0 data we choose to treat the two datasets in the same way when averaging. 
Assuming LTE, the ratio of the integrated intensities can be used to measure the excitation temperature \citep{Goldsmith99}:
\begin{equation}
\frac{\nu_u^2 A_l I_u}{\nu_l^2 A_u I_l} = \frac{g_u}{g_l} \mathrm{e}^{-E_{ul}/k T}
\end{equation}
where \textit{$\nu$} is the line frequency, \textit{A} is the Einstein coefficient, \textit{I} is the integrated intensity, \textit{g} is the statistical weight for a linear rotor, \textit{E$_{ul}$} is the energy difference between the two transitions, \textit{k} is the Boltzmann constant, and \textit{T} is the excitation temperature. 
Integrating the ring-averaged spectrum for each transition gives an integrated intensity (and one sigma uncertainty) of $5.1\pm0.2$ K km s$^{-1}$ for the 1-0 transition and $28.7\pm0.8$ K km s$^{-1}$ for the 4-3 transition. Using these values we find an excitation temperature of 39~K, with a one sigma uncertainty of 2~K. 

\section{Discussion}
At the radii we consider, 0.8-2\arcsec, the radial temperature profile based on observations of \coo\ is roughly constant at 21~K \citep{Schwarz16}. That the temperature probed by \coo\ remains constant over many radii suggests that the \coo\ emission is originating primarily from just above the CO snow surface. This can also be seen for the more inclined IM Lup disk, where the CO snow surface is directly imaged in addition to constraining the CO freeze out temperature to $\sim21$~K \citep{Pinte18a}.
As such the 21~K from \coo\ provides an upper limit for the midplane temperature at these radii. This is consistent with models of the TW Hya disk \citep[e.g.,][] {Du15,Kama16b}, which set the midplane temperature in this region at temperatures between 10-20~K. Thus, the $39\pm2$~K gas where \nnhp\ is emitting resides above the midplane.

In the disk models of \citet{Aikawa15} and \citet{vantHoff17} (itself based on the \citet{Kama16b} model), which focus specifically on the \nnhp\ 4-3 emission, the 40~K gas temperature contour is at a scale height of $z/r\approx0.2$ for the radii where \nnhp\ emission is observed. 
Both sets of models predict \nnhp emission at this scale height. In the \citet{Aikawa15} models including millimeter grains, CO has been converted to \cotwo\ ice at a scale height of 0.2 while photodissociation prevents \nn\ from being reprocessed into \ammonia\ ice. This combination of a low CO abundance and a high \nn\ abundance results in a layer of \nnhp. \cotwo\ ice has been proposed as a potential reservoir of volatile carbon in disks such as TW Hya with a low CO gas abundance \citep{Eistrup16,Bosman18,Schwarz18}. It is also worth noting that our derived temperature for the \nnhp\ emitting layer of $39\pm2$~K is close to the expected desorption temperature of \cotwo\ ice.

Alternatively, in the \citet{vantHoff17} models a surface layer of \nnhp\ is generated when CO has been dissociated by UV photons while \nn\ remains self-shielded.
These models are specifically tailored to TW Hya while considering only a small network of chemical reactions.
The best fit to the observed \nnhp\ 3-2 emission in TW Hya occurs when both the CO and \nn\ gas abundances have been reduced, with a total \nn/CO ratio of 1. 
In summary, there are a variety of factors which influence the morphology of the \nnhp\ emission in TW Hya, with photodissociation and the CO and \nn\ gas abundances being of particular importance. 
While several combinations of processes match the observed emission, it is clear that in this system \nnhp\ emission is not a good tracer of the CO snowline deeper in the disk. 
That the \nnhp\ emission in TW Hya originates from a surface layer was also suggested by \citet{Nomura16} based on the observed brightness temperature of the 4-3 line.

\section{Summary}
We use averaged observations of the \nnhp\ J = 1-0 and J = 4-3 lines from $0\farcs8$ to $2\arcsec$ in TW Hya to derive the temperature of the \nnhp\ emitting layer. 
We find an excitation temperature of 39~K with a one sigma uncertainty of 2~K, significantly warmer than the expected midplane temperature of $<$ 20~K at the radii where \nnhp\ is observed to emit. 
Therefore we conclude that in TW Hya \nnhp\ primarily emits from a surface layer, with the vertical boundaries set by processes such as photodissociation or chemical reprocessing, rather than a layer deeper in the disk bounded by the direct freeze-out of \nn\ and CO. 
These results highlight the importance of understanding protoplanetary disk structure when interpreting molecular line observations.

\acknowledgments
This paper makes use of the following ALMA data: JAO.ALMA\#2011.0.00340.S and JAO.ALMA \#2016.0.00592.S.
ALMA is a partnership of European
Southern Observatory (ESO) (representing its member
states), National Science Foundation (USA), and National
Institutes of Natural Sciences (Japan), together with National
Research Council (Canada), National Science Council and
Academia Sinica Institute of Astronomy and Astrophysics
(Taiwan), and Korea Astronomy and Space Science Institute
(Korea), in cooperation with Chile. The Joint ALMA
Observatory is operated by ESO, Associated Universities,
Inc/National Radio Astronomy Observatory (NRAO), and
National Astronomical Observatory of Japan. The National
Radio Astronomy Observatory is a facility of the National
Science Foundation operated under cooperative agreement by
Associated Universities, Inc.
This work was supported by funding from NSF grant
AST-1514670 and NASA NNX16AB48G. 
K.S. acknowledges the support of NASA through Hubble Fellowship Program grant HST-HF2-51419.001, awarded by the Space Telescope Science Institute, which is operated by the Association of Universities for Research in Astronomy, Inc., for NASA, under contract NAS5-26555.

\facility{ALMA}
\software{CASA v4.7.0 \citep{casa}, eddy \citep{eddy}, matplotlib \citep{matplotlib}, numpy \citep{numpy}}


\end{document}